\documentclass{icrc}

\usepackage{times}
\usepackage{graphicx} 

\newbox\grsign \setbox\grsign=\hbox{$>$} \newdimen\grdimen
\grdimen=\ht\grsign
\newbox\simlessbox \newbox\simgreatbox
\setbox\simgreatbox=\hbox{\raise.5ex\hbox{$>$}\llap
     {\lower.5ex\hbox{$\sim$}}}\ht1=\grdimen\dp1=0pt
\setbox\simlessbox=\hbox{\raise.5ex\hbox{$<$}\llap
     {\lower.5ex\hbox{$\sim$}}}\ht2=\grdimen\dp2=0pt
\def\simgreat{\mathrel{\copy\simgreatbox}}
\def\simless{\mathrel{\copy\simlessbox}}
\newbox\simppropto
\setbox\simppropto=\hbox{\raise.5ex\hbox{$\sim$}\llap
     {\lower.5ex\hbox{$\propto$}}}\ht2=\grdimen\dp2=0pt

\begin{document}

\sloppy

\title{The Status of the STACEE Observatory}

\author[1]{C.E. Covault}
\affil[1]{Department of Physics,
Case Western Reserve University, Cleveland, OH 44106, USA}
\author[2]{L.M. Boone}
\affil[2]{Santa Cruz institute for Particle Physics, 
University of California, Santa Cruz, CA 95064, USA}
\author[3]{D. Bramel}
\affil[3]{Columbia University \& Barnard College, New York, NY 10027, USA}
\author[4]{E. Chae}
\affil[4]{Enrico Fermi Institute, University of Chicago, 5640 Ellis Av., Chicago, IL 60637, USA}
\author[5]{P. Fortin}
\affil[5]{Department of Physics, McGill 
University, Montreal, Quebec H3A 2T8, Canada}
\author[6,7]{D.M Gingrich}
\affil[6]{Centre for Subatomic Research,
University of Alberta, Edmonton, Alberta T6G 2N5, Canada}
\affil[7]{TRIUMF, Vancouver, British Columbia V6T 2A3, Canada}
\author[4]{J.A. Hinton}
\author[5]{D.S. Hanna}
\author[3]{R. Mukherjee}
\author[5]{C. Mueller}
\author[8]{R.A. Ong}
\affil[8]{Department of Physics \& Astronomy,
University of California, Los Angeles, CA 90095, USA}
\author[5]{K. Ragan}
\author[4]{R.A. Scalzo}
\author[8]{D.R. Schuette}
\author[5]{C.G. Theoret}
\author[2]{D.A. Williams}

\correspondence{covault@hep.uchicago.edu}

\firstpage{1}
\pubyear{2001}

\maketitle

\begin{abstract}
The Solar Tower Atmospheric Cherenkov Effect Experiment (STACEE)
is a ground-based instrument designed to study astrophysical
sources of gamma radiation in the energy range of 50 to 500 GeV.
STACEE uses an array of large heliostat mirrors at the National
Solar Thermal Test Facility in Albuquerque, New Mexico, USA.
The heliostats are used to collect Cherenkov light produced
in $\gamma$-ray air showers.  The light is concentrated onto an
array of photomultiplier tubes located near the top of a tower.
The construction of STACEE started in 1997 and has been completed
in 2001.  During the 1998-99 observing season, we used a portion
of the experiment, STACEE-32, to detect $\gamma$-rays from the
Crab Nebula.  The completed version of STACEE uses 64 heliostat
mirrors, having a total collection area of 2300 m$^{2}$.
During the last year, we have also installed custom electronics
for pulse delay and triggering, and 1 GHz Flash ADCs to read out
the photomultiplier tubes.  The commissioning of the full
STACEE instrument is underway.  Preliminary observations and
simulation work indicate that STACEE will have an energy
threshold below 70 GeV and a reach for extragalactic $\gamma$-ray
sources out to redshift of $\sim$1.0. In this
paper we describe the design and performance of STACEE.
\end{abstract}

\section{Overview}

STACEE (The Solar Tower Atmospheric Effect Experiment) is a new
experiment constructed for ground-based gamma-ray astrophysics. STACEE 
detects $\gamma$-rays via the atmospheric Cherenkov
radiation produced in extensive air showers. STACEE uses 64 large
solar mirrors (heliostats) as the primary collector of Cherenkov
photons.  This very large collection area allows STACEE to probe a
region of $\gamma$-ray energies (50-500 GeV) not yet explored by previous
imaging Cherenkov instruments.

STACEE is located at the National Solar Thermal Test Facility (NSTTF)
at Sandia National Laboratories in Albuquerque, New Mexico, USA (see
Figure~1).  STACEE is one of four solar-tower Cherenkov experiments
that are currently in operation or under development.  These include
the French CELESTE experiment in the Pyrenees \citep{celeste} the
Spanish/German GRAAL experiment \citep{graal} and the Solar Two
Gamma-Ray Observatory near Barstow, California \citep{solar2}.

\begin{figure}
 \includegraphics[width=8.3cm]{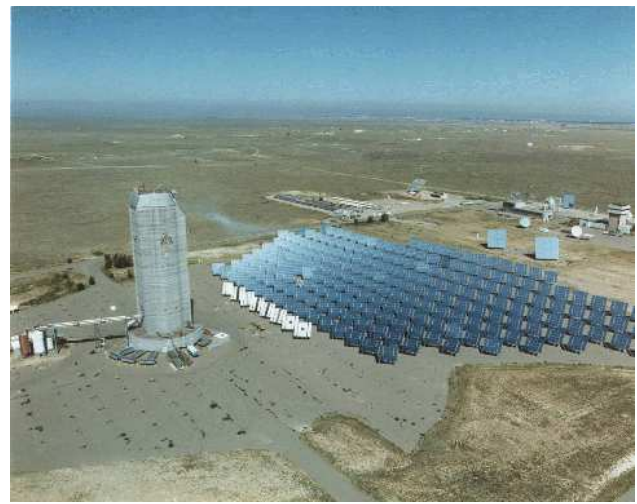}
 \caption{Aerial view of National Solar Tower Test Facility
 at Sandia National Laboratories in Albuquerque, New Mexico,
 USA.}
\end{figure}

The STACEE experiment has been built in stages, with each stage using
progressively more heliostats and progressively improved optics and
electronics all of which are incorporated in the final STACEE
experiment that is now complete.  Figure~2 shows a plan view of the
heliostats at the NSTTF used for each stage of the STACEE experiment.
The stages of construction are delineated as follows:

\begin{figure}
 \includegraphics[width=8.3cm]{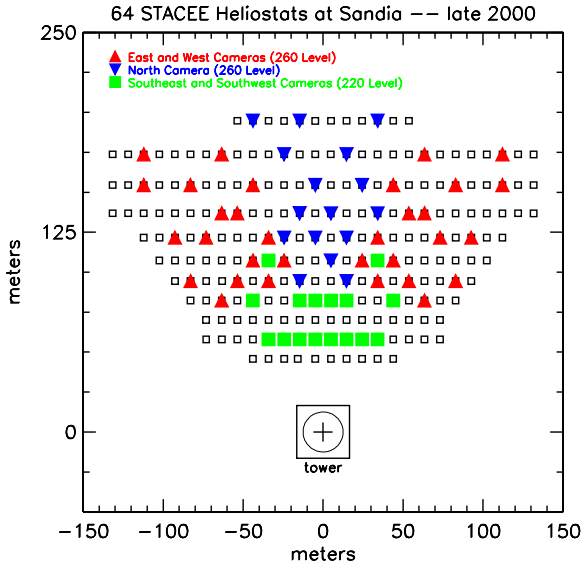} 
 \caption{Plan view of NSTTF showing heliostats
used for STACEE.  STACEE-32 used the East and West
Cameras; STACEE-48 used the East, West, and North Cameras.}
\end{figure}

\begin{description}
\item{{\bf STACEE-32:}} Construction on STACEE began in 1997.  By October of
1998 we commissioned a partially complete system with two secondary
mirrors and 32 heliostats called STACEE-32.  During the 1998-1999
observing seasons, we operated STACEE-32 to measure instrument
performance and to conduct initial observations of gamma-ray
sources. These results have been reported
elsewhere \citep{rene_snowbird,corbin_snowbird}.  In
particular, we observed the Crab nebula and pulsar between
December 1998 and March 1999 with STACEE-32.  We detected 
the Crab with high statistical
significance (seven standard deviations) at an energy of threshold of 
190 $\pm$ 60 GeV.  This
detection represents the first astrophysical results from STACEE 
(Oser, 2001) and demonstrated the viability of the instrument
concept.
\item{{\bf STACEE-48:}} A second stage of the partially completed
STACEE instrument was constructed during 1999 and 2000, with 48
heliostat channels active.  In addition, several key improvements were
made to the optics and electronics (see below).  STACEE-48 has been
operating as a gamma-ray observatory since November 2000.  Sources
observed thus far with STACEE-48 include Markarian 421, which has been
detected by STACEE with high significance \citep{jim_icrc}, and
Markarian 501 \citep{rene_icrc}.
\item{{\bf STACEE:}} Construction of the full STACEE instrument
with 64 heliostat channels and five secondary mirrors
is now complete.   Regular
operation of the completed STACEE experiment 
as a gamma-ray observatory is expected to commence
by Fall, 2001.  At this point STACEE will operate continuously
for several years \citep{rene_icrc}.
\end{description}

\section{Completion of the STACEE instrument}
During the last two years, we have worked on the completion of
the STACEE instrument.  We carried out the following tasks:

\begin{itemize}
\item We have learned that the alignment of the heliostat facets is
   an important consideration for the overall optical throughput.
   To obtain the best throughput possible, we have developed a
   precise algorithm for setting facet headings using lasers
   and a telescope.  Individual facet headings are now
   determined with an accuracy corresponding to $\simless$20 cm
   of movement of the heliostat spot at the position of the
   secondary mirror.

\item A set of alignment lasers has been mounted at key positions
   on the optical structures on the tower to verify and monitor
   the optical alignment of secondary mirrors and cameras.
   Individual PMT positions are now verified monthly using a
   CCD camera and moonlight.  We have also used moonlight to
   directly determine the proper bias (offset) heading for each
   heliostat to obtain maximum light throughput.

\item To reduce background light due to reflected albedo, the asphalt
   field beneath the heliostats was coated by a dark black sealant.
   This coating reduced the albedo by up to 50\%.

\item An infrared cloud monitor and an optical photometer (which
   tracks a star close to the STACEE field of view) are being
   commissioned to work along with STACEE.
\end{itemize}

During the last two years, we have also made a push to install the
final STACEE electronics. This work included the installation and
integration of additional trigger electronics to support all 64 PMT
channels.  We have designed and installed a custom dynamic delay and
trigger system, based in VME using FPGA technology \citep{mcgill}.
The system features delays programmable in 1-nsec steps over a
one-microsecond range and a two-level multiplicity trigger with an
adjustable trigger coincidence gate (currently set at 12 nsec). The
trigger system keeps a record of hits on each channel for the 32 nsec
window surrounding a trigger, with a 1 nsec accuracy.

We have also completed the installation of high speed (1 GHz) 8-bit
FADCs to digitize signals from the PMT channels. These
state-of-the-art digitizers (Acqiris DC270) provide the accurate time
and amplitude measurements necessary for complete shower
reconstruction.  Improved shower reconstruction will help STACEE
achieve a better angular resolution and gamma/hadron separation (and
hence sensitivity).

Figure~3 shows a schematic of the electronics for STACEE. The FADCs
measure the analog pulse of each PMT channel to determine its
time-of-arrival and amplitude.  They also measure the instantaneous
night sky background before and after the Cherenkov signals.  The
FADCs serve as an analog pipeline to buffer the PMT information while
the trigger decision is being made.  Conventional multi-hit TDCs are
used for verification of the pulse timing.

The FADCs used by STACEE are packaged in a 4-channel compact-PCI
(cPCI) board, and are controlled using a real-time Linux-based DAQ
that operates in conjunction with the VME-based trigger system.
STACEE personnel developed and implemented the entire suite of
software required for the DAQ, including the kernel-based device
driver; this work was necessary because the base drivers written for
Windows NT would have unacceptable dead-time.  A full crate of FADCs
to provide readout for 48 PMT channels has been installed, and the
commissioning tests indicate that we will be able to readout the FADCs
with a very small dead-time ($\sim$2 ms/event, as compared to
$\sim$10 ms/event
for the VME portion of the DAQ).  The raw data taken by the FADCs is
fully integrated into the standard STACEE analysis chain, and during
the summer we will finalize the reconstruction algorithms to provide
robust estimates of the pulse amplitude and time.

\begin{figure}
 \includegraphics[width=8.3cm]{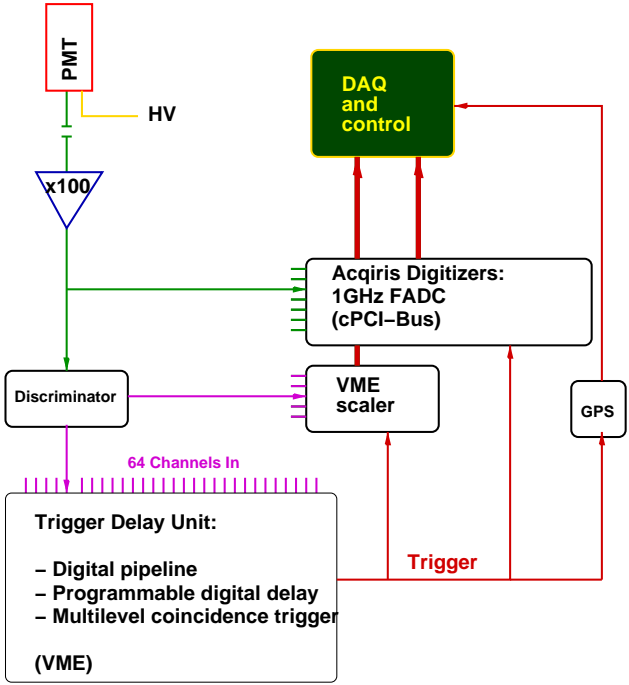}
 \caption{Schematic of STACEE electronics, including the new
delay/trigger circuitry and the Acqiris digitizers.}
\end{figure}

\section{Recent performance results from STACEE-48}

A concerted effort is underway to characterize the detector
performance.  These studies are based upon {\em in situ} calibration, Monte
Carlo simulation, and the analysis of data taken.  Event
reconstruction parameters, in particular, angular reconstruction and
energy determination, are primarily determined from the arrival times
and pulse amplitudes of the Cherenkov signal seen at each heliostat.
Thus, calibrating and characterizing STACEE's timing resolution and
optical response is critical.  Here we present some of the results
that have been obtained thus far in this effort.

We have developed an initial reconstruction scheme that is applied to
the STACEE Cherenkov data.  To reconstruct the arrival direction of an
incoming shower, the times recorded by the TDCs (after calibration and
trims are applied) are fit to a spherical shape.  Figure~4 shows a
plot of the timing residuals on one channel after best fit spherical
wavefront reconstruction.  The residual width of $\sim$1 ns is at the level
predicted by simulation indicating that the timing measurement from
any one channel is primarily limited by the intrinsic fluctuations in
the shower. 

\begin{figure}
 \includegraphics[width=8.3cm]{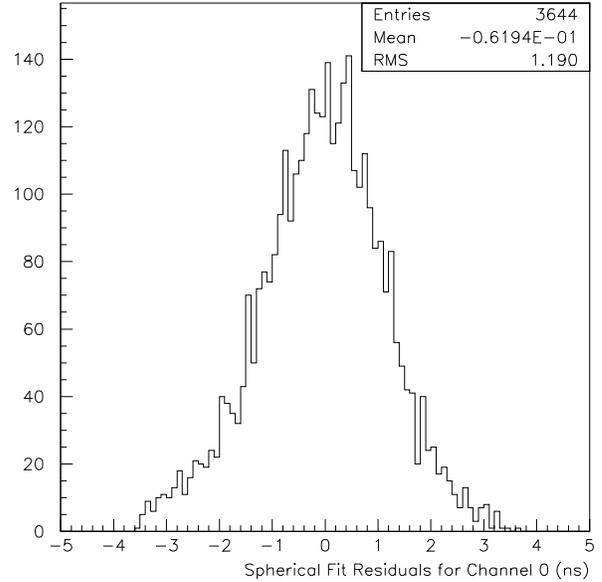}
 \caption{Timing residuals on a single STACEE channel relative to
  best-fit spherical wavefront reconstruction.
  The RMS of this distribution is 1.2 ns, indicating that
  STACEE is achieving excellent time resolution.}
\end{figure}

We obtain an estimate of the pointing accuracy of STACEE by dividing
the array into two overlapping sub-arrays and using each sub-array to
reconstruct the arrival direction separately.  The angular difference
between the reconstructed directions from the sub-arrays, as shown in
Figure~5, is a measure of the intrinsic angular resolution (modulo a
statistical correction factor).  From this study we determine an
angular resolution of $\simless$0.2$^o$, which agrees with simulation and
the design value.  Figure~5 also shows the reduced $\chi^{2}$
distribution for the goodness-of-fit to a spherical wavefront.

\begin{figure}
 \includegraphics[width=8.3cm]{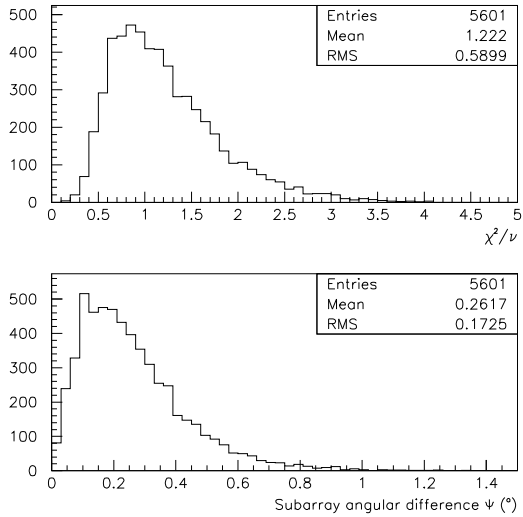}
 \caption{{\bf Top:} Distribution of reduced $\chi^{2}$
   values from the best fit spherical 
   wave front to arrival times recorded at each heliostat.
   {\bf Bottom} Split-array
   reconstructed direction difference, which is a good estimate of
   angular position resolution on reconstructed showers.}
\end{figure}

Another key indicator of detector performance is the trigger rate
versus discriminator threshold curve which provides a good measure of
the effective operating threshold of the experiment.  Figure~6 shows
this curve for STACEE-48 observing cosmic rays coming from the
zenith.  Using this data, we can set the operating threshold at the
lowest possible value that ensures a negligible contribution to the
rate from accidental triggers on night-sky noise.  We note that
for STACEE-48, the usable threshold is constrained by the
intrinsic limit of 16 hits per channel on multi-hit TDCs and by
deadtime associated with individual discriminators. The recent
installation of FADCs will allow us to run STACEE at the lowest
possible trigger threshold, which translates into the lowest possible
energy threshold.

\begin{figure}
 \includegraphics[width=8.3cm]{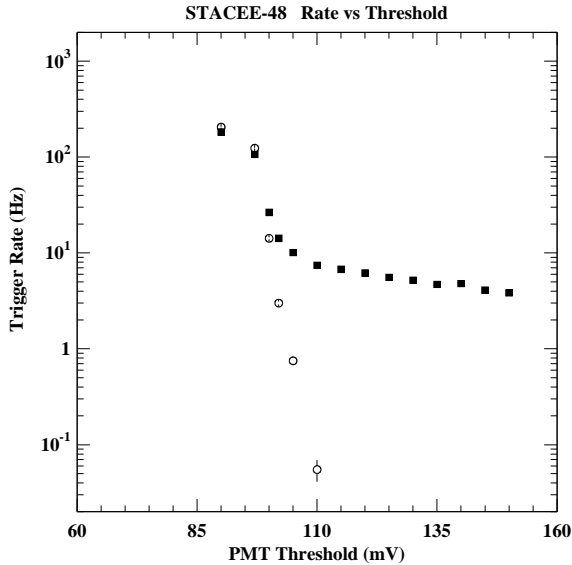} 
 \caption{Trigger rate
 versus discriminator threshold for STACEE-48. The black squares
 indicate the response of the experiment with channel delays set
 in-time for Cherenkov triggers.  The open circles, taken with
 delays times scrambled, represent the rate due to accidental triggers
 on the night sky background. The turnover at rates above 100 Hz is
 due to DAQ deadtime.  Above a threshold of 120 mV, the
 experiment operates with Cherenkov rate $\simgreat$7.5 Hz and with
 negligible accidental background.  For comparison, one
 photo-electron corresponds to $\sim$25 mV.}
\end{figure}

Finally, work is in progress to compare in detail results from
simulations against data collected by the experiment to determine
energy threshold.  We use the CORSIKA package to simulate air showers
\citep{corsika}.  Figure~7 shows the preliminary energy response of
the STACEE-48 system to simulated $\gamma$-rays for data taken
Winter/Spring 2001 giving a threshold of $120 \pm 25$ GeV assuming
a source spectrum of the form  $E^{-2}$ located at zenith. With the
implementation of FADCs for STACEE we expect to operate at a lower
threshold ($\simless$100 GeV).

\begin{figure}
 \includegraphics[width=8.3cm]{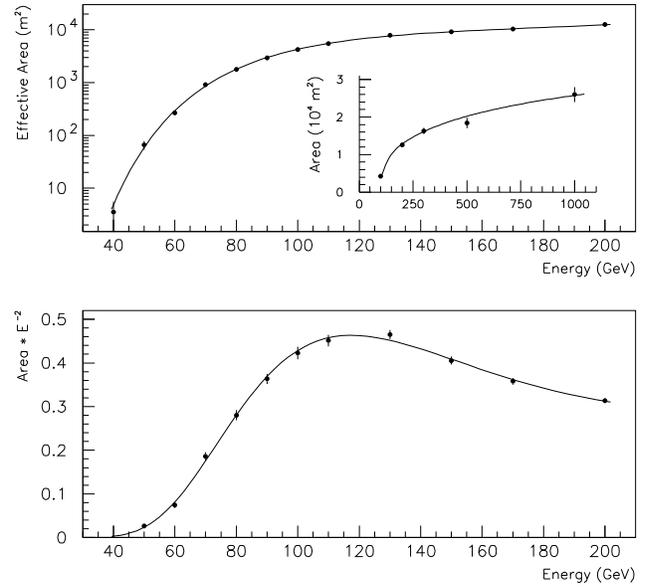}
 \caption{The STACEE-48 energy response to
 $\gamma$-rays based on Monte Carlo simulations. Our preliminary
 estimate of the energy 
 threshold is $120 \pm 25$ GeV at zenith 
 for observations made in Winter/Spring 2001. This energy threshold
 is appropriate for observations made with a 145~mV PMT threshold,
 well above the threshold where the rate of accidental triggers 
 becomes neglibible.}
\end{figure}

\begin{acknowledgements}
We are grateful to the staff at the NSTTF for their excellent support.
Thanks to Gora Mohanty, Jeff Zweerink, Tumay Tumer, Marta Lewandowska,
Scott Oser, and Fran\c{c}ois Vincent.  This work was supported in part
by the National Science Foundation (under Grant Numbers PHY-9983836,
PHY-0070927, and PHY-007095), the Natural Sciences and
Engineering Research Council, FCAR (Fonds pour la Formation de
Chercheurs et l'Aide \`a la Recherche), the Research Corporation, and
the California Space Institute.  CEC is a Cottrell Scholar of Research
Corporation.
\end{acknowledgements}

\end{document}